\documentclass[aps,prd,reprint,showpage,showpacs,nobibnotes,superscriptaddress,twocolumn,amssymb,amsmath,nofootinbib,floatfix]{revtex4-1}
\usepackage{graphicx}
\usepackage[utf8x]{inputenc}
\usepackage{color}
\usepackage{subfig}
\usepackage{multirow}
\usepackage{slashed}
\usepackage{ulem}

\begin{document}

\title{Left-right symmetric extensions of 3-3-1 models }


\author{Edison T. Franco}%
\email{edisonfranco@uft.edu.br}
\affiliation{
Universidade Federal do Tocantins - Campus Universit\'ario
de Aragua\'\i na\\
Av. Paraguai (esquina com Urixamas), Cimba \\ Aragua\'\i na - TO, 77824-838, Brazil
}

\author{V. Pleitez}%
\email{vicente@ift.unesp.br}
\affiliation{
Instituto  de F\'\i sica Te\'orica--Universidade Estadual Paulista \\
R. Dr. Bento Teobaldo Ferraz 271, Barra Funda\\ S\~ao Paulo - SP, 01140-070,
Brazil
}

\date{02/21/2017
%
}
\begin{abstract}
We propose four left-right symmetric extensions of chiral 3-3-1 models. Although they have some common features they also have important differences due to different representation content.
\end{abstract}


\maketitle

\section{Introduction}
In the electroweak standard model (ESM),  left-handed fermions transform in a different way than the right-handed components. Hence, it is a chiral model.  In this way parity violation is accommodated but not explained. The parity issue was the motivation of the early left-right symmetric extensions of the ESM~\cite{Mohapatra:1974gc,Senjanovic:1975rk,Senjanovic:1978ev}.
In this sort of models the electroweak gauge symmetry is extended to $SU(2)_L\otimes SU(2)_R\otimes U(1)_{B-L}$~\cite{Mohapatra:1980qe} and, besides the explanation of parity violation as a spontaneously broken symmetry, the model is able to generate Dirac~\cite{Senjanovic:1978ev} or Majorana ~\cite{Mohapatra:1979ia,Mohapatra:1980yp} masses for neutrinos. A left-right symmetric extension can be implemented for all chiral models.

Here, we will consider that sort of extensions for models with 3-3-1 chiral symmetry in such a way that the gauge symmetries are
$G_{3331}\equiv SU(3)_C\otimes SU(3)_L\otimes SU(3)_R\otimes U(1)_X$ with the  general
charge operator defined as $Q=T^L_3+T^R_3+\beta(T^L_8+T^R_8)+X$, specifically with $\beta=-\sqrt{3}$ (models Ia, Ib) and $\beta=-1/\sqrt{3}$ (models IIa, IIb). For a general classification of 3-3-1 chiral models with arbitrary $\beta$ see Ref.~\cite{Diaz:2004fs}.

In all these extensions the anomalies are canceled considering three generations only as in the chiral 3-3-1 version since left- and right- fermion triplets have the same $U(1)_X$-charge. A model similar with our model IIa was proposed in Ref.~\cite{Dias:2010vt} but those authors used only scalar triplets and for this reason fermion masses are generated only by five-dimensional operators.  Here we will use in both sort of models only renormalizable interactions. In fact, the first time that this sort of model was briefly discussed was in Ref.~\cite{Pleitez:1994pu}.

In Secs.~\ref{sec:modelsI} and we introduce the models Ia and Ib, while in Sec.~\ref{sec:modelsII} we consider the models IIa and IIb.
In Sec. \ref{sec:discussion} we discuss the phenomenology concerning to these types of models.
Finally, our conclusions and outlook are given in Sec. \ref{sec:conclusions}.

\section{Models I}\label{sec:modelsI}
In the model Ia, the particle content is \cite{Pisano:1991ee,Frampton:1992wt,Foot:1992rh}:  left-handed quarks  $Q_{aL}=(d^\prime_a,-u^\prime_a,j^\prime_a)^T_L\sim (\textbf{3},\textbf{3}^*_L,\textbf{1}_R,-1/3)$,
$Q_{3L}=(u^\prime_3,d^\prime_3,J^\prime)^T_L\sim (\textbf{3},\textbf{3}_L,\textbf{1}_R,2/3)$; and the
right-handed quarks: $Q_{aR}=(d^\prime_a,-u^\prime_a,j^\prime_a,)^T_R\sim (\textbf{3},\textbf{1}_L,\textbf{3}^*_R,-1/3)$,
$Q_{3R}=(u^\prime_3,J^\prime, d^\prime_3)^T_R\sim (\textbf{3},\textbf{1}_L,\textbf{3}_R,2/3)$. In the lepton sector we have
left-handed leptons: $\psi_{lL}=(\nu^\prime_l,l^\prime,l^{\prime c})^T_L\sim (\textbf{1},\textbf{3}_L,\textbf{1}_R,0)$; and
right-handed leptons: $\psi_{lR}=(\nu^\prime_l,E^\prime _l,E^{\prime c}_l)^T_R\sim (\textbf{1},\textbf{1}_L,\textbf{3}_R,0)$.
Primed fields denote symmetry eigenstates. The left-handed and right-handed fields are not equivalent since in each triplet the charged degrees of freedom are complete. Only neutrino has one component in each sector. At the energy scale at which the symmetry is valid all fields are uncharged and massless and this does not matter.

In the scalar sector we introduce the following multiplets: $T\sim(\textbf{1},\textbf{3}_L,\textbf{3}^*_R,0)$, 
$P\sim(\textbf{1},\textbf{3}_L,\textbf{3}_R,1)$, which couple to quarks, $T$ also couples with leptons but the sextets,
$S_L\sim(\textbf{1},\textbf{6}^*_L,\textbf{1}_R,0)$ and
$S_R\sim(\textbf{1},\textbf{1}_L,\textbf{6}^*_R,0)$, couple only with leptons.
The electric charge assignment of the
bi-triplets $T$, $P$ and the sextets $S_{L,R}$ are as follows:
\begin{equation}
T=\left(
\begin{array}{ccc}
t_{1}^{0} & t_{1}^{-} & t_{2}^{+} \\
t_{3}^{+} & t_{2}^{0} & t_{1}^{++} \\
t_{4}^{-} & t_{2}^{--} & t_{3}^{0}%
\end{array}%
\right) ,\quad P=\left(
\begin{array}{ccc}
\rho _{1}^{+} & \rho _{1}^{0} & \rho _{1}^{++} \\
\rho _{2}^{0} & \rho _{2}^{-} & \rho _{3}^{+} \\
\rho _{2}^{++} & \rho _{4}^{+} & \rho ^{+++}%
\end{array}%
\right) ,
\end{equation}
\begin{equation}
S_{L,R}=\left(
\begin{array}{ccc}
s_{1L,R}^{0} & \frac{s_{1L,R}^{+}}{\sqrt{2}} & \frac{s_{2L,R}^{-}}{\sqrt{2}}
\\
\frac{s_{1L,R}^{+}}{\sqrt{2}} & s_{1L,R}^{++} & \frac{s_{2L,R}^{0}}{\sqrt{2}}
\\
\frac{s_{2L,R}^{-}}{\sqrt{2}} & \frac{s_{2L,R}^{0}}{\sqrt{2}} & s_{2L,R}^{--}%
\end{array}%
\right) ,
\end{equation}%
Notice that $P$ involves a scalar Higgs with triple charge. We define
\begin{equation}
 \left\langle T\right\rangle =\left(
\begin{array}{ccc}
v_{d} & 0 & 0 \\
0 & v_{u} & 0 \\
0 & 0 & v_{j}%
\end{array}%
\right), \quad
\left\langle P\right\rangle =\left(
\begin{array}{ccc}
0 & V_{d} & 0 \\
V_{u} & 0 & 0 \\
0 & 0 & 0%
\end{array}%
\right)
\end{equation}
\begin{equation}
\left\langle S_{L,R}\right\rangle =\left(
\begin{array}{ccc}
v_{L,R1} & 0 & 0 \\
0 & 0 & \frac{v_{L,R2}}{\sqrt{2}} \\
0 & \frac{v_{L,R2}}{\sqrt{2}} & 0%
\end{array}%
\right).
\end{equation}
Because of the bi-triplet $P$, the symmetry is breakdown to $U(1)_Q$. Here we will assume, for the sake of simplicity, that all VEVs are real.

The Yukawa interactions in the quark sector are:
\begin{eqnarray}
-\mathcal{L}^Y_q&=&\bar{Q}_{aL}G_{ab}Q_{bR}T^\dagger
+\bar{Q}_{3L}G_{3a}Q_{aR}P \nonumber \\
&+&\bar{Q}_{aL}G_{a3}Q_{3R}P^\dagger
+\bar{Q}_{3L}G_{33}Q_{3R}T
+ H.c.,
\label{yukaquarks1}
\end{eqnarray}
$a,b=1,2$ and 3 are generation indices, and we have omitted $SU(3)$ indices and summation over flavors. Notice that with only the bi-triplet $T$ only the first two families mix, $G_{ab}$ is a $2\times2$ matrix, and $G_{33}$ is a constant. Besides, $\langle T\rangle\not=0$ leaves an extra $U(1)$ unbroken symmetry. All of these is corrected by adding a second bi-triplet $P$ carrying $U(1)_X$-charge.

The Yukawa interactions in the lepton sector are:
\begin{eqnarray}
-\mathcal{L}_{l}^{Y}&=&\bar{\psi}_{iL}G_{ij}^{l}\psi
_{jR}T+\frac{1}{2}\overline{(\psi _{iL})^{c}}G^{S}_{ij}\psi _{jL}S_{L}\nonumber \\ &+&\frac{1}{2}\overline{(\psi
_{iR})^{c}}G^{S}_{ij}\psi _{jR}S_{R}+H.c.\label{yukaleptons1}
\end{eqnarray}%
where summation over flavours $i,j=1,2,3$ has been omitted and $G^S$ is a symmetric matrix.

We add a parity $\mathcal{P}$ as follows:
\begin{eqnarray}
&& g_L\leftrightarrow  g_R,\; Q_{aL}\leftrightarrow Q_{aR}, \; Q_{3L}\leftrightarrow Q_{3R}, \;\psi_L \leftrightarrow \psi_R,\nonumber \\ &&
T\leftrightarrow T^\dagger , \;  P\leftrightarrow P, \;S_L\leftrightarrow S_R, \; W_L\leftrightarrow W_R\;,
\label{parity}
\end{eqnarray}
which implies that $G=G^{\dagger}$ in Eq.~(\ref{yukaquarks1}), $G^{l}=G^{l \dagger}$ in Eq.~(\ref{yukaleptons1}) and thus all mass matrices are Hermitian.

In the quark sector we have, for type-$u$ quarks in the basis
$(-u_1^\prime, -u_2^\prime, u_3^\prime)$ and in the type-$d$ sector in the basis
$(d_1^\prime, d_2^\prime, d_3^\prime)$ the respective mass matrices are given by
\begin{equation}
M^{u}=\left(
\begin{array}{ccc}
G_{11}v_{u} & G_{12}v_{u} & G_{13}V_{u} \\
G_{12}^{\ast}v_{u} & G_{22}v_{u} & G_{23}V_{u} \\
G_{13}^{\ast} V_{u} & G_{23}^{\ast}V_{u} & G_{33}v_{d}%
\end{array}
\right) ,
\end{equation}
\begin{equation}
 M^{d}=\left(
\begin{array}{ccc}
G_{11}v_{d} & G_{12}v_{d} & G_{13}V_{d} \\
G_{12}^{\ast}v_{d} & G_{22}v_{d} & G_{23}V_{d} \\
G_{13}^{\ast}V_{d} & G_{23}^{\ast}V_{d} & G_{33}v_{u}
\end{array}
\right).
\end{equation}

Since the mass matrices are symmetric ($M^{u,d}=(M^{u,d})^{\dagger}$) they are diagonalized by one unitary matrix and not by two unitary matrices. The exotic quarks have their masses given by $M^{j}=Gv_{j}$ and $M^{J}=G_{33}v_{j}$, where $G$ is a $2\times2$ matrix and $G_{33}$ is a constant.

In the lepton sector we have that leptons with the same electric are separated in two $6\times6$  matrices. In the lepton sector, in the basis $\ell^\prime_{L,R} = (l^{\prime }, E^{\prime})^T_{L,R}$, we have
\begin{equation}\label{massl1charged}
M^{l}=\left(
\begin{array}{cc}
G^S\frac{v_{2L}}{\sqrt{2}} & G^{l}v_{u}
\\
G^{lT}v_{j} &G^S\frac{v_{2R}}{\sqrt{2}}
\end{array}
\right)_{6 \times 6},
\end{equation}
and in the neutrinos sector we have
\begin{equation}\label{massl1neutrinos}
 M^{\nu }=\left(
\begin{array}{cc}
G^{S}v_{1L} & G^{l}v_{d} \\
G^{lT}v_{d} & G^{S}v_{1R}
\end{array}
\right)_{6 \times 6},
\end{equation}\noindent
where the mass term for the neutrinos is written in the form $-\mathcal{L}_{\nu }^{mass}=\frac{1}{2}\overline{n_{L}}M^{\nu }\left( n_{L}\right)
^{c}+H.c.$ in the basis $n_{L}=(\nu_{L}^{\prime } , \nu _{R}^{\prime c})^T$, and
where $G^{S,l}$ are arbitrary $3\times3$ hermitian matrices. If $v_{R2}\gg v_{L2},v_u,v_j$ and $v_{R1}\gg v_{L1},v_d$, the exotic leptons are heavier than the known leptons,
and the right-handed neutrinos are heavier than the left-handed (active) ones. In fact, in this model there are both, type-I and type-II, seesaw mechanisms and the neutrino mass matrix in Eq.~(\ref{massl1neutrinos}) is  $6\times6$.

Notice that if the sextet are not added or if $\langle S_{L,R}\rangle=0$, the known  charged leptons gain mass only through their mixing with the exotic ones $E$. Moreover in this case the charged leptons and the neutrino matrices are diagonalized by the same transformation, thus the PMNS matrix must be trivial. Hence, the sextets are necessary to obtain the correct lepton masses and the PMNS mixing matrix.

In the present case neutrinos are Majorana particles but if, instead of the sextets $S_L$ and $S_R$ we introduces triplets, we would have Dirac neutrinos. For instance $\eta_L\sim(\textbf{1},\textbf{3},\textbf{1},0),\eta_R \sim(\textbf{1},\textbf{1},\textbf{3},0)$, the Yukawa interactions $G^{\eta}L\psi^T_L\epsilon C\psi_L\eta_L$ and $G^{\eta}\psi^T_R\epsilon C  \psi_R\eta_R$ are allowed, but the masses obtained are not still realistic ones because $G^{\eta}$ is $3\times3$ antisymmetric matrix.

A different model (model Ib), with the same quark content as in model Ia and also with $\beta=-\sqrt{3}$, can be obtained if $l^c$ is substituted by a new charged lepton $E^+$ that may be considered as a particle instead of an antiparticle. See Ref.~\cite{Pleitez:1992xh}.
In this case $\psi_{lL}=(\nu^\prime,l^\prime,E^{+\prime}_l)^T_L\sim(\textbf{3},\textbf{1},0)$ and $\psi_{lR}=(\nu^\prime,l^\prime,E^{+\prime}_l)^T_R\sim(\textbf{1},\textbf{3},0)$, and all the lepton masses are generated by the Yukawa interactions like Eq.~(\ref{yukaleptons1}). This model was also discussed in Sec.~II of Ref.~\cite{Huong:2016kpa}.

The charged lepton mass matrix is now given by
\begin{equation}
M^{l}=\left(
\begin{array}{cc}
G^{l}v_{u} & G^S \frac{v_{2L}}{\sqrt{2}} \\
 G^S \frac{v_{2R}}{\sqrt{2}} &
G^{lT}v_{j}
\end{array}
\right)_{6 \times 6},
\label{massl1b}
\end{equation}
also in the basis $\ell^\prime_{L,R} = (l^{\prime }, E^{\prime})^T_{L,R}$. The neutrino and quark sectors are the same as in the previous model. Yet, if the sextets are not introduced the exotic charged leptons decouple from the light ones. However, as in the previous model, if the sextets are not introduced the mass matrices of charged leptons and neutrinos are proportional, thus inducing a trivial PMNS.

The scalar potential in models I is given by
\begin{widetext}
\begin{eqnarray}
V &=&\mu _{T}^{2}\text{Tr}T^{\dagger }T+\mu _{P}^{2}\text{Tr}P^{\dagger
}P+\mu _{S}^{2}\text{Tr}(S_{L}^{\ast }S_{L}+S_{R}^{\ast }S_{R})+\lambda _{1}%
\text{Tr}(T^{\dagger }T)^{2}+\lambda _{2}(\text{Tr}(T^{\dagger }T))^{2}
+\lambda _{3}\text{Tr}(P^{\dagger }P)^{2}\nonumber\\
&+&\lambda _{4}(\text{Tr}%
(P^{\dagger }P))^{2}\notag +\lambda _{5}\text{Tr}(S_{L}^{\ast }S_{L}+S_{R}^{\ast
}S_{R})^{2}+\lambda _{6}(\text{Tr}(S_{L}^{\ast }S_{L}+S_{R}^{\ast
}S_{R}))^{2} +\text{Tr}[(S_{L}S_{L}^{\ast }+S_{R}S_{R}^{\ast })(\lambda _{7}T^{\dagger
}T+\lambda _{8}P^{\dagger }P)] \notag \\
&+&\text{Tr}[(S_{L}^{\ast }S_{L}+S_{R}^{\ast
}S_{R})]\text{Tr}[\lambda _{9}(T^{\dagger }T)+\lambda _{10}(P^{\dagger }P)]
+\lambda _{11}[\text{Tr}(T^{\dagger }TP^{\dagger }P)+\text{Tr}(TT^{\dagger
}P^{\dagger }P)]+\lambda _{12}\text{Tr}(T^{\dagger }T)\text{Tr}(P^{\dagger
}P)  \notag \\
&+&\lambda _{13}[\text{Tr}(P^{\dagger }S_{L}^{\ast })\text{Tr}(PS_{L})+\text{%
Tr}(P^{\dagger }S_{R}^{\ast })\text{Tr}(PS_{R})]
+\lambda _{14}[\mathrm{Tr}\left( T^{\dagger }T P^{T}P^{\ast }\right)+\mathrm{Tr}\left( T T^{\dagger } P^{T}P^{\ast }\right)]  \notag \\
&+&\lambda _{15}[(S_{R}^{\ast
}T)(S_{L}T)+H.c.]+f_{1}[(TTT)+H.c.]+f_{2}[(S_{L}S_{L}S_{L})+H.c.]+f_{3}[(S_{R}S_{R}S_{R}) + H.c.].
\label{potential1}
\end{eqnarray}
\end{widetext}

\section{Models II}
\label{sec:modelsII}
In model IIa we have the following representation content:  left-handed quarks  $Q_{aL}=(d^\prime_a,-u^\prime_a,D^\prime_a)^T_L\sim (\textbf{3},\textbf{3}^*_L,\textbf{1}_R,0)$,
$Q_{3L}=(u^\prime_3,d^\prime_3,U^\prime)^T_L\sim (\textbf{3},\textbf{3}_L,\textbf{1}_R,1/3)$; and the
right-handed quarks: $Q_{aR}=(d^\prime_a,-u^\prime,D^\prime)^T_R\sim (\textbf{3},\textbf{1}_L,\textbf{3}^*_R,0)$,
$Q_{3R}=(u^\prime_3,d^\prime_3,U^\prime)^T_R\sim (\textbf{3},\textbf{1}_L,\textbf{3}_R,1/3)$. In the lepton sector we have
left-handed leptons: $\psi_{lL}=(\nu^\prime_l,l^\prime,N^\prime_l)^T_L\sim (\textbf{1},\textbf{3}_L,\textbf{1}_R,-1/3)$; and
right-handed leptons: $\psi_{lR}=(\nu^\prime_l,l^\prime,N^\prime_l)^T_R\sim (\textbf{1},\textbf{1}_L,\textbf{3}_R,-1/3)$.
Primed fields denote symmetry eigenstates. In model IIa, $N^\prime_L$ and $N^\prime_R$ are new fields, as in the model of Refs.~\cite{Singer:1980sw,Valle:1983dk}. However, if $N^\prime_{lL}$ is identified with $(\nu_{lR})^c$ as in the models of Refs.~\cite{Montero:1992jk,Foot:1994ym}. In models II, the scalar multiplets are: $T^\prime\sim(\textbf{1},\textbf{3}_L,\textbf{3}^*_R,0)$, and $H\sim (\textbf{1},\textbf{3}_L,\textbf{3}_R,1/3)$, defined as follows
\begin{equation}
T^{\prime }=\left(
\begin{array}{ccc}
\tilde{t}_{1}^{0} & \tilde{t}_{1}^{-} & \tilde{t}_{2}^{0} \\
\tilde{t}_{2}^{+} & \tilde{t}_{3}^{0} & \tilde{t}_{3}^{+} \\
\tilde{t}_{4}^{0} & \tilde{t}_{4}^{-} & \tilde{t}_{5}^{0}%
\end{array}%
\right) ,~H=\left(
\begin{array}{ccc}
h_{1}^{+} & h_{1}^{0} & h_{2}^{+} \\
h_{2}^{0} & h_{3}^{-} & h_{3}^{0} \\
h_{4}^{+} & h_{4}^{0} & h_{5}^{+}%
\end{array}%
\right).
\end{equation}
We define the VEVs in the following form:
\begin{equation}
\left\langle T^{\prime }\right\rangle =\left(
\begin{array}{ccc}
v_{d} & 0 & v_{R} \\
0 & v_{u} & 0 \\
v_{L} & 0 & v_{D}%
\end{array}%
\right) ,~\left\langle H\right\rangle =\left(
\begin{array}{ccc}
0 & V_{u} & 0 \\
V_{d} & 0 & V_{D} \\
0 & V_{U} & 0%
\end{array}%
\right).
\end{equation}

The Yukawa interactions are given by
\begin{eqnarray}
-\mathcal{L}^Y_q&=&\bar{Q}_{aL}G_{ab}Q_{bR}T^{\prime \dagger}
+\bar{Q}_{3L}{G}_{3a}Q_{aR}H\nonumber \\ &+&\bar{Q}_{aL}G_{a3}Q_{3R}H^\dagger+ \bar{Q}_{3L}G_{33}Q_{3R}T^\prime
+ H.c.,
\label{yukaquarks2}
\end{eqnarray}
in the quark sector and
\begin{eqnarray}
-\mathcal{L}^Y_l&=&\overline{(\psi_{iL})}G^l_{ij}\psi_{jR}T^\prime
+ H.c.,
\label{yukaleptons2}
\end{eqnarray}
in the lepton sector.

The parity $\mathcal{P}$ in this case is
\begin{eqnarray}
&& g_L\leftrightarrow  g_R,\; Q_{aL}\leftrightarrow Q_{aR}, \; Q_{3L}\leftrightarrow Q_{3R}, \;\psi_L \leftrightarrow \psi_R, \; \nonumber\\&&T^\prime\leftrightarrow T^{^\prime \dagger} , \;  H\leftrightarrow H, \; W_L\leftrightarrow W_R\;,
\label{parity2}
\end{eqnarray}
and again, all the $G$ matrices are hermitian.

The mass matrix in the quark sector are:
\begin{equation}
M^u=\left(
\begin{array}{cccc}
v_{u}G_{11} & v_{u}G_{12} & V_{u}G_{13} & V_{U}G_{13} \\
v_{u}G_{12}^{\ast } & v_{u}G_{22} & V_{u}G_{23} & V_{U}G_{23} \\
V_{u}G_{13}^{\ast } & V_{u}G_{23}^{\ast } & v_{d}G_{33} & v_{R}G_{33} \\
V_{U}G_{13}^{\ast } & V_{U}G_{23}^{\ast } & v_{L}G_{33} & v_{D}G_{33}%
\end{array}%
\right),
\end{equation}%
for the $u$-type in the basis $(-u_1^\prime, -u_2^\prime, u_3^\prime,U^\prime)$,  and
\begin{equation}
M^d=\left(
\begin{array}{ccccc}
v_{d}G_{11} & v_{d}G_{12} & V_{d}G_{13} & v_{L}G_{11} & v_{L}G_{12} \\
v_{d}G_{12}^{\ast } & v_{d}G_{22} & V_{d}G_{23} & v_{L}G_{12}^{\ast } &
v_{L}G_{22} \\
V_{d}G_{13}^{\ast } & V_{d}G_{23}^{\ast } & v_{u}G_{33} & V_{D}G_{13}^{\ast }
& V_{D}G_{23}^{\ast } \\
v_{R}G_{11} & v_{R}G_{12} & V_{D}G_{13} & v_{D}G_{11} & v_{D}G_{12} \\
v_{R}G_{12}^{\ast } & v_{R}G_{22} & V_{D}G_{23} & v_{D}G_{12}^{\ast } &
v_{D}G_{22}%
\end{array}%
\right),
\end{equation}
for the $d$-type in the basis $(d_1^\prime, d_2^\prime, d_3^\prime, D_1^\prime, D_2^\prime)$. Notice that if we set $v_{L,R}=V_{U,D}=0$ the known quarks do not couple with the new ones $U^\prime$ and $D^\prime_{1,2}$, however it is better assume only that   $v_x/v\ll1$, where $v_x=v_{L,R},V_{U,D}$ and $v$ is a VEV related to the scale at which left-right symmetry is manifested. This is because if there is no mixing among all quarks of the same charge, an automatic $\mathbb{Z}_2$ survive and the extra quarks are stable, or at least long lived, unless the  interactions with the scalars mix all the singly charge scalars.
The masses in the lepton sector are given by
\begin{equation}\label{massl2charged}
M^{l}=G^{l}v_{u},
\end{equation}
\begin{equation}\label{massl2neutrinos}
\quad M^{\nu }=\left(
\begin{array}{cc}
0 & M_{D} \\
M_{D}^{T} & 0%
\end{array}%
\right)_{12 \times 12},
\end{equation}
where
\begin{equation}
\label{massl2MD}
M_{D}=\left(
\begin{array}{cc}
G^{l}v_{d} & G^{l}v_{R} \\
G^{l}v_{L} & G^{l}v_{D}%
\end{array}%
\right)_{6 \times 6},
\end{equation}
in the basis
$n_{L}^{T}=(\nu _{L}^{\prime } , N_{L}^{\prime } , \nu _{R}^{\prime c} , N_{R}^{\prime c})^T$.
In this model the neutrinos become Dirac particles.
It is clear to see from Eqs.~(\ref{massl2neutrinos}) and ~(\ref{massl2MD}) that the light neutrinos ($\nu$) decouple from the heavy states ($N$) in the mass matrix only if  $v_L=v_R=0$, but this case is not realistic since the PMNS matrix is also the identity matrix.

In model IIb we have the same quark representations as in model IIa but in the lepton sector we have the following representation content: $\psi_{lL}=(\nu^\prime_l,l^\prime,{\nu^\prime_{l}}^{c})^T_L\sim (\textbf{1},\textbf{3}_L,\textbf{1}_R,-1/3)$; and
right-handed leptons: $\psi_{lR}=(N^\prime_l,l^\prime,{N^\prime_{l}}^{c})^T_R\sim (\textbf{1},\textbf{1}_L,\textbf{3}_R,-1/3)$.
The neutrino mass matrix, $M^\nu$, have the same structure as given by Eqs.~(\ref{massl2neutrinos}) and~(\ref{massl2MD}), but in 
the basis $n_{L}^{T}=(\nu _{L}^{\prime }, N_{L}^{\prime }  , N_{R}^{\prime c}, \nu _{R}^{\prime c}) ^{T}$. This shows that in the model IIb all the mass terms mixes the light neutrinos with heavy states, thus there is no decoupling.

\begin{widetext}
The scalar potential in these case is given by
\begin{eqnarray}
V &=&\mu _{T^{\prime }}^{2}\mathrm{Tr}\left( T^{\prime \dagger }T^{\prime
}\right) +\mu _{H}^{2}\mathrm{Tr}\left( H^{\dagger }H\right) +\lambda _{1}%
\mathrm{Tr}\left( T^{\prime \dagger }T^{\prime }\right) ^{2}+\lambda
_{2}\left( \mathrm{Tr}\left( T^{\prime \dagger }T^{\prime }\right) \right)
^{2}+\lambda _{3}\mathrm{Tr}\left( H^{\dagger }H\right) ^{2}  \nonumber \\
&+&\lambda
_{4}\left( \mathrm{Tr}\left( H^{\dagger }H\right) \right) ^{2} +\lambda _{5}[\mathrm{Tr}\left( T^{\prime \dagger }T^{\prime }H^{\dagger
}H\right)+\mathrm{Tr}\left(T^{\prime }T^{\prime \dagger }H^{\dagger
}H\right)]+\lambda _{6}\mathrm{Tr}\left( T^{\prime \dagger }T^{\prime }\right) \mathrm{Tr}\left(
H^{\dagger }H\right)  \nonumber \\
&+& \lambda _{7}[\mathrm{Tr}\left( T^{\prime \dagger }T^{\prime
}H^{T}H^{\ast }\right)+\mathrm{Tr}\left( T^{\prime}T^{\prime \dagger }H^{T}H^{\ast }\right)] +f_{1}\left( \mathrm{Tr}\left( T^{\prime }T^{\prime
}T^{\prime }\right) +H.c.\right).\label{potential2}
\end{eqnarray}
\end{widetext}

\section{Discussion}\label{sec:discussion}
All 3-3-1 models have interesting new processes that deserve to be searched in accelerators. In particular the minimal model ($\beta=-\sqrt{3}$), predicts new resonances formed by exotic quarks in processes like
\begin{eqnarray}
&& pp(J\bar{J}u\bar{u}V^-V^+)\to e^-_R e^+_L+\nu^c_L \nu_R (\slashed{E})+jets,\nonumber \\ &&
pp(J\bar{J}d\bar{u}V^-U^{++})\to e^+_Le^-_Re^-_Le^+_R+\nu^c_L(\slashed{E})+jets,\nonumber \\&&
pp(j\bar{j}d\bar{d}V^-V^+)\to e^+_R e^-_L+\nu_L\nu_L(\slashed{E})+jets,\nonumber \\ &&
pp(j\bar{j}u\bar{u}U^{--}U^{++})\to l^-l^+l^-l^++jets,
\label{processe331}
\end{eqnarray}
which are interesting because the resonances involving the exotic quarks are long-lived, since they are protected by a $\mathbb{Z}_2$ symmetry which is broken by scalar interactions. In the left-right symmetric extensions this model has  interactions with like those in Eq.~(\ref{yukaquarks1})  with the bi-triplet $P$ and quarks: $G_{a3}\overline{J}[d^\prime_a \rho^{++}+u^\prime_3\rho^++j_{a}^{\prime }
\rho ^{+++}]$, this is an example of interactions in which we can recognize that the exotic particles $J,j,\rho^{+++,++,+}$ are produced by pairs since they are odd under $\mathbb{Z}_2$ but $u,d$-type quarks are even. In this case the exotic particles will be stable.
Fortunately, we have verified that all singly charged scalars mix in the mass matrices and that in the scalar potential there are interactions of exotic scalars and those coupled with known quarks and leptons as for instance that in the $\lambda_{13}$ in Eq.~(\ref{potential1}),
$\rho ^{+++}s_{2L}^{--}s^-_{1L}  \rho _{1}^{0*}(\rho _{2}^{0*})$,
and from the Yukawa interaction in Eq.~(\ref{yukaleptons1}), we see that there are interactions like
$G_{ij}^{S}[\overline{( l_{iL}^{\prime }) ^{c}}s_{1L}^{+}\nu _{jL}^\prime+\overline{l_{iR}^{\prime }}s_{2L}^{--}( l_{jR}^{\prime
}) ^{c}] $
allowing the decay of $\rho^{+++}$ to leptons through $s_{1L}^{+}$ and $s_{2L}^{++}$. Hence, in particular in models I, there is a resonance with triple electric charge $\Delta^{+++}(\bar{j} J)$ which can decay into the scalar $\rho^{+++}$ and, since there are interactions like those in the $\lambda_{13}$ term discussed above, the decay  $\rho^{+++}\to s_1^{+}s_2^{++}$ does occurs, after $\rho _{1}^{0}$ and/or $\rho _{2}^{0}$ get VEV, and $\rho ^{+++}\to s_{1L}^{+}s_{2L}^{++}\to
l_{R}^{+}l_{R}^{+}l_{L}^{+}\nu _{L}^{c}$ is allowed. Hence, the decay $\Delta^{+++}\to 3l^++3\nu$ may happen. All these models have 17 vector bosons, in particular the $\beta=-\sqrt{3}$ have doubly charge vector bileptons $U^{++}_{L,R}$ which could have interesting phenomenological consequences at colliders~\cite{Cao:2016uur}.

It is well known that chiral 3-3-1 models have a Landau-like pole at a particular value of $s^2_W$. In the chiral models with $\beta=-\sqrt{3}$, the pole occurs when $s^2_W=0.25$, at an energy scale of a few TeVs~\cite{Dias:2004dc,Dias:2006ns}, while in models with $\beta=-1/\sqrt{3}$, the pole occurs when $s^2_W=0.75$, and the energy scale may be larger than the grand unification scale. In the respective left-right symmetric extensions, the pole occurs when
$s^2_W=1/8$ and $s^2_W=3/8$ in model of type I and type II, respectively~\cite{Dias:2010vt}. On the other hand, at the $Z$-pole, $s^2_W(M_Z)\approx0.231$ hence, in type I models $s^2_W(pole)<s^2_W(M_Z)$.
Hence, at first sight these models are excluded.  However, this is not necessarily the case. It is still possible that in these sort of models $s^2_W$, at higher energies, run to values that are lower than the value at the $Z$-pole.  The running of $s^2_W$ depends on the representation content of the models. In fact, this behavior occurs in other models. For instance, in left-right symmetric $SU(5)\otimes SU(5)$, $s^2_W$ decreases from low to high energies, essentially due to the scalar and fermion triplets~\cite{EmmanuelCosta:2011wp}, which also are present in the type I models. This can be seen as a prediction of the these models: if in the future at high energies $s^2_W$ only increases with the energy, this will certainly rule out these sort of models.
The energy scale of the Landau pole has not been calculated yet in both I and II models. We note however, that because of the scalar bi-triplets, needed to generate all fermions masses with only renormalizable interactions, there is a mixing among all vector ($W_L,W_R$ and $Z_L,Z_R$) and scalar bosons of the same charge or $C\!P$ properties. Then, the energy scale of the poles may change. This important issue will be considered elsewhere.

Moreover, it is worthy to note the following. As a consequence of the quark representation content, the chiral models with $\beta=-\sqrt3$ are not embedded straightforward in a grand unified theory (GUT). Due to the existence in those models of extra quarks with large $U(1)_X$ hypercharge, more quarks with appropriate value of $X$ have to be added in order to satisfy $\textrm{Tr}X=0$ in GUT multiplets. This is not necessarily a fault of the models, it may be a virtue. For instance, since the Landau pole occurs at relatively low energies (about 4 TeV), the electroweak scale can be stabilized, i.e, there is no hierarchy problem~\cite{Dias:2009au} and also the model allows a dynamically symmetry breaking to the SM also at the scale of a few TeVs with all the extra gauge bosons and the exotic quarks acquiring masses much larger than the scale of electroweak symmetry breaking~\cite{Das:1999hn}. And after all, who said that grand unification of three of the fundamental interactions should be straightforward? However, for a possible unification for models with $\beta=-1/\sqrt{3}$ see \cite{Byakti:2013uya} and references therein.

\section{Conclusions}\label{sec:conclusions}
Here we have considered four left-right symmetric extensions of chiral 3-3-1 models with only renormalizable interactions. The models I introduces exotic quarks and the models II only extends the quarks sector for four (type-u quarks) or five (type-d quarks) generations with heavy neutrinos.

For each model we have considered all possible breaking of symmetry allowing all VEV to be nonvanishing. In this vein, we have shown that the models of type I leads to a seesaw mechanism in the lepton sector, essentially driven by the introduction of left-right sextets. while the models of type II induces a mixing in the quark matrices and the sextets are not needed anymore. The mixing of all scalar sector allows a decay signature of the $\rho^{+++}$ into light fermions, thus the consistence of the models can be tested at colliders. On the other hand, to be consistent the models I should allow the decreasing $s^2_W$ with increasing of energy. This will be addressed elsewhere.

After we have finished this work we saw that in Refs.~\cite{Reig:2016tuk,Dong:2016sat} similar left-right symmetric extensions of 3-3-1 models were considered. Those authors consider a general analysis of the models in term of the $\beta$ parameter in the definition of the electric charge operator. In our case we have considered $\beta=-\sqrt{3},-1/\sqrt{3}$ each one with two different representation content. It is worthing to note that some characteristics do not depend on the value of $\beta$. For instance, in the model with $\beta=-\sqrt{3}$ we have two possibilities in the lepton sector: i) $\psi_l=(\nu_l,l,l^c)^T_L,\;\psi_l=(\nu_l,E,E^c)^T_R$; or, ii) $\psi_l=(\nu_l,l^-,E^+)^T_L,\;\psi_l=(\nu_l,l^-,E^+)^T_R$. This difference is not trivial: in the first  case the flavor lepton numbers is violated because the charged lepton and their charge conjugate are in the same triplet while in the second case we can assign to $E^+$ the same flavor lepton number of the known charged leptons. Moreover the neutral current couplings are different in both models with the same $\beta$. Furthermore, another difference is evident comparing the matrix in Eq.~(\ref{massl1neutrinos}) (model Ia) with those in Eqs.~(\ref{massl2neutrinos}) and~(\ref{massl2MD}) (model IIa), in particular neutrinos are Majorana in the former case, and Dirac in the second one.
Moreover in the case of Eqs.~(\ref{massl2neutrinos}) and ~(\ref{massl2MD}) 
a type-I Dirac seesaw mechanism is implemented. Models with $SU(3)_C\otimes SU(M)_L\otimes SU(N)_R\otimes U(1)_X$ has been considered recently in Ref.~\cite{Dong:2016sat}. We would like to stress that phenomenology independently of the value of $\beta$ can be made only in some simple situation as, for instance, considering only the $Z^\prime$ of the models~\cite{Buras:2014yna}.

\acknowledgments{ ETF thanks to IFT-UNESP for its kind hospitality where part of this work was done. VP thanks to CNPq for partial financial support.}

\end{document}